\documentclass[11pt,a4paper]{article}

\usepackage{amsmath,amssymb}
\usepackage{epsfig,graphicx}
\usepackage{subfigure}
\usepackage{graphicx}
\usepackage{rotating}
\usepackage{cancel}
\usepackage{bm}
\usepackage{color}
\usepackage{comment}
\usepackage{cite}
\usepackage{axodraw4j}

\renewcommand\({\left(}
\renewcommand\){\right)}
\renewcommand\[{\left[}

\newcommand{\exclude}[1]{}

\def\beq{\begin{equation}}
\def\eeq{\end{equation}}

\begin{document}
\numberwithin{equation}{section}
\title{{\normalsize  \mbox{}\hfill ~}\\
\vspace{2.5cm} 
\Large{\textbf{A Family of WISPy Dark Matter Candidates
\vspace{0.5cm}}}}

\author{Joerg Jaeckel\\[2ex]
\small{\em Institut f\"ur theoretische Physik, Universit\"at Heidelberg, Philosophenweg 16, 69120 Heidelberg, Germany}\\[0.5ex]
}

\date{}
\maketitle

\begin{abstract}
Dark matter made from non-thermally produced bosons can have very low, possibly sub-eV masses. Axions and hidden photons are prominent examples of such ``dark'' very weakly interacting light (slim) particles (WISPs). A suitable mechanism for their non-thermal production is the misalignment mechanism. Their dominant interaction with Standard Model (SM) particles is via photons. 
In this note we want to go beyond these standard examples and discuss a wide range of scalar and pseudo-scalar bosons interacting with SM matter fermions via derivative interactions. Suitably light candidates arise naturally as pseudo-Nambu-Goldstone bosons. In particular we are interested in examples, inspired by familons, whose interactions have a non-trivial flavor structure.
\noindent
\end{abstract}

\newpage

\section{Introduction}
Despite many years of study the nature of dark matter (DM) is still shrouded in a veil of darkness. We basically only know that something exists, it is dark and it clumps, i.e. it forms structures.
If DM is made out of particles this means that they
have to be sufficiently long-lived, dark (i.e. very weakly interacting with photons\footnote{Observations like, e.g., the bullett cluster also suggest that the self-interaction cross section of DM is also quite small.}) and they have to be very slowly moving in today's Universe.
In absence of more detailed information a plethora of candidates have been suggested axions and WIMPs being perhaps the most prominent~\cite{Bertone:2004pz,Sikivie:2006ni,Jaeckel:2010ni}.

So why add another one? As long as we have not yet discovered a dark matter particle (and we certainly have tried)
it is important to ask where it could hide. From a phenomenological point of view it is therefore important
to try to cover dark matter candidates which have widely different properties and therefore may require very different  detection techniques. One regime that so far has received less attention is that of very low masses. Notable exceptions are axions and hidden photons\footnote{Of course, one could include neutrinos in the list, but at least standard left handed neutrinos are too hot to be a large fraction of DM.}~\cite{Preskill:1982cy,Abbott:1982af,Dine:1982ah,Nelson:2011sf,Arias:2012az}. For direct detection purposes their most relevant interaction with the SM is via photons. So in this note we go into an orthogonal direction and ask: what if the dominant interaction with the SM is via couplings to fermions.

In particular we focus on derivative type interactions as they are typical for (pseudo-)Nambu-Goldstone bosons. While we ask this question in general we consider in particular interactions with non-trivial flavor structure. The classic example being (pseudo-)familons~\cite{Wilczek:1982rv}.
In part we are motivated in this by seeking interesting and unusual signals. The other part is the remarkable property that, as we will see later, for light (pseudo-)scalars interactions with non-trivial flavor structure are often less constrained than those of first generation particles.

In specific models pseudo-familons as a dark matter candidate have been discussed in~\cite{Carone:2012dg,Joshipura:1987tf,Berezhiani:1989fp}. Aside from attempting a more general and less model-dependent discussion our work differs in two crucial aspects.
First, we use the non-thermal misalignment mechanism for the production of the dark matter particles. In contrast to~\cite{Carone:2012dg} this allows us to have dark matter masses which are very small possibly in the sub-eV range.
Second, the specific models~\cite{Joshipura:1987tf,Berezhiani:1989fp} identify the dark matter particle also with the axion. While this particle can have flavor changing couplings it also has the typical flavor independent axion couplings, which severely constrain it. As mentioned above,  interactions with non-trivial flavor structure are often significantly less constrained, and our analysis exhibits large regions in parameter space suitable for dark matter.

The note is structured as follows. In the following Sect.~\ref{goldstone} we will briefly review pseudo-Nambu-Goldstone bosons and their interactions with matter. Next in Sect.~\ref{misalignment} we will point out the crucial features of the misalignment mechanism for the non-thermal production of dark matter. In Sects.~\ref{diagonal} we will then discuss flavor diagonal and non-diagonal couplings. We will conclude in Sect.~\ref{conclusions}.

\section{Pseudo-Nambu-Goldstone bosons and their interactions with matter}\label{goldstone}
Let us briefly recall how Goldstone bosons arise and how they interact with matter (see also~\cite{Feng:1997tn}).
\subsection*{Goldstone bosons and their interactions}
In general Goldstone bosons arise when continuous global symmetries are broken.
Their interactions with other particles are via derivative couplings,
\begin{equation}
\frac{1}{f^{a}_{X}}\partial_{\mu}\phi^{a}J^{\mu,a},
\end{equation}
where $\phi_{a}$ denotes the Goldstone bosons corresponding to a symmetry transformation and $J^{\mu,a}$ is the Noether current for this symmetry transformation.
$f^{a}_{X}$ is the Goldstone decay constant,
\begin{equation}
f^{a}_{X}=\sqrt{2 
\langle F^{\dagger}\rangle T^{a}_{F}T^{a}_{F} \langle F\rangle
},
\end{equation}
where $F$ is the field that spontaneously breaks the symmetry, $T_{F}$ are the generators implementing the symmetry transformation on $F$, and $\langle F\rangle$ the vacuum expectation value.

For simplicity in the following we will restrict ourselves to the case of a single symmetry transformation, effectively a U(1) symmetry, and drop the index $a$.

We are interested in family symmetries, acting on the fermions of the Standard Model. For concreteness and simplicity we restrict ourselves to symmetries in the lepton sector. But generalization to the quarks is fairly straightforward.
Let us take the symmetry transformations,
\begin{equation}
L^{a}_{L}\rightarrow U_{L}^{ab}L^{b}_{L},\quad E^{a}_{R}\rightarrow U_{E}^{ab}E^{b}_{R},
\end{equation}
with $3\times 3$ matrices $U$,
\begin{equation}
U_{L}=\exp\left(i\alpha T_{L}\right),\quad U_{E}=\exp\left(i\alpha T_{E}\right).
\end{equation}
Moreover, $L_{L}$ denotes the left handed lepton doublets and $E_{R}$ the right handed charged fermion fields. $a,b$ are generation indices.
The generators $T_{L},T_{E}$ give the specifics of the transformation.
For example,
\begin{equation}
T_{L}=
\left(\begin{array}{ccc}
1 & 0 & 0\\
0 & 0 & 0\\
0 & 0 & 0 \\
\end{array}\right),\qquad T_{E}=\left(\begin{array}{ccc}
1 & 0 & 0\\
0 & 0 & 0\\
0 & 0 & 0 \\
\end{array}\right)
\end{equation}
corresponds to a first generation lepton family number U(1) symmetry\footnote{In general for it to be a true U(1) symmetry one needs to ensure that there is a finite $\alpha\neq 0$ such that $U_{L}=U_{E}=1$. In our case this is obviously the case for $\alpha=2\pi$. If this is not the case we have a non-compact symmetry $\mathbb{R}$. Such a situation may arise if one considers non-trivial directions of larger symmetry groups.}.
The conserved current in this case is easily found to be,
\begin{equation}
J^{\mu}=\bar{e}\gamma^{\mu}e+\bar{\nu}_{e}\gamma^{\mu}\nu_{e},
\end{equation}
corresponding to the (electron+electron neutrino) number current and a conserved first generation lepton number.

Returning to the more general case. For the charged lepton fields we can write the interaction in terms of Dirac spinors $E$ (containing both left and right handed components,
\begin{eqnarray}
\label{goldinteraction}
{\mathcal{L}}_{\rm int}\!\!&=&\!\!\frac{\partial_{\mu}\phi(x)}{f_{X}}\left[\bar{E}\frac{T_{E}+T_{L}}{2}\gamma^{\mu}E+\bar{E}\frac{T_{E}-T_{L}}{2}\gamma^{\mu}\gamma^{5}E\right]
\\\nonumber
\!\!&=&\!\!\frac{\partial_{\mu}\phi(x)}{f_{X}}\left[\bar{E}T_{+}\gamma^{\mu}E+\bar{E}T_{-}\gamma^{\mu}\gamma^{5}E\right],
\end{eqnarray}
where in the last line we have grouped the interaction into scalar and pseudo-scalar parts (both can be present!).

For tree-level processes we can use the Dirac equation to simplify the interaction,
\begin{eqnarray}
\label{effectivegoldstone}
{\mathcal{L}}_{\rm int}\!\!&=&\!\!-i\frac{\phi(x)}{f_{X}} \left[\bar{E}\left[T_{+}M_{E}-M_{E}T_{+}\right]E-\bar{E}\left[T_{-}M_{E}+M_{E}T_{-}\right]\gamma^{5}e\right]
\\\nonumber
\!\!&=&\!\!-i\frac{\phi(x)}{f_{X}} \left[\bar{E}\left[T_{+},M_{E}\right]E-\bar{E}\left\{T_{-},M_{E}\right\}\gamma^{5}E\right].
\end{eqnarray}
Here $M_{E}$ is the charged lepton mass matrix which we have taken to be hermitean.

Importantly we note, that these effective interactions are proportional to the masses of the particles involved. Therefore interactions with neutrinos are typically orders of magnitude weaker. 

Let us consider two examples. The first is a generalisation of the one briefly mentioned above, with the symmetry only acting on the first generation 
\begin{equation}
\label{example1}
T_{+}=
a\left(\begin{array}{ccc}
1 & 0 & 0\\
0 & 0 & 0\\
0 & 0 & 0 \\
\end{array}\right),\qquad 
T_{-}=b\left(\begin{array}{ccc}
1 & 0 & 0\\
0 & 0 & 0\\
0 & 0 & 0 \\
\end{array}\right).
\end{equation}
(indeed our simple example above corresponds to $a=1,\,b=0$.
Inserting this into Eq.~\eqref{effectivegoldstone} we find for the effective interaction of the electrons,
\begin{equation}
\label{example12}
{\mathcal{L}}_{\rm int} = i\frac{2m_{e}b}{f_{X}}\phi \bar{e}\gamma^{5} e.
\end{equation}
Importantly we note that the scalar part vanishes and only the pseudo-scalar part survives. Accordingly the scalar interactions are strongly suppressed in this case.

Let us now consider a second example with a slightly more non-trivial family symmetry exchanging electrons and muons,
\begin{equation}
\label{mix}
T_{+}=
a\left(\begin{array}{ccc}
0 & 1 & 0\\
1 & 0 & 0\\
0 & 0 & 0 \\
\end{array}\right),\qquad 
T_{-}=b\left(\begin{array}{ccc}
0 & 1 & 0\\
1 & 0 & 0\\
0 & 0 & 0 \\
\end{array}\right).
\end{equation}
In this case the effective interaction reads,
\begin{eqnarray}
\label{mux}
{\mathcal{L}}_{\rm int}\!\!&=&\!\!a\frac{m_{\mu}-m_{e}}{f_{X}}\phi\bar{\mu}e+b\frac{m_{\mu}+m_{e}}{f_{X}}\phi\bar{\mu}\gamma^{5}e
+h.c.
\\\nonumber
\!\!&\approx &\!\! \frac{m_{\mu}}{f_{X}}\left[a \phi\bar{\mu}e+b\phi\bar{\mu}\gamma^{5}e\right]
+h.c. .
\end{eqnarray}

This can serve as a simple prototype for a Goldstone interaction with a non-trivial flavor structure.

\subsection*{Massive pseudo-Goldstone bosons}
Exact Goldstone bosons are, of course, massless and as such unsuitable to being dark matter.
Therefore we are more interested in massive pseudo-Goldstone bosons.

To do this we can simply introduce a small explicit breaking of the symmetry in question.

Let us briefly recall for the case of a U(1) symmetry that this gives us a periodic potential in $\phi$ with a periodicity of $f_{X}$.
Normalizing the smallest charge to $1$ all symmetry transformations are 
\begin{equation}
T_{X}=\exp(iq_{X}\alpha),
\end{equation}
where $q_{X}$ is the integer charge of the transformed field.
For $\alpha=2\pi$ we are then effectively back in the original state.

We can now use that the Goldstone field is directly linked to a symmetry transformation
\begin{equation}
\alpha\equiv \frac{\phi}{f_{X}}.
\end{equation}
It is clear that $\phi$ has a periodicity,
\begin{equation}
\phi\rightarrow \phi+2\pi f_{X}.
\end{equation}
This restricts the potential to be
\begin{equation}
\label{goldpot}
V(\phi)=\sum_{n} V_{n} \cos\left(n\frac{\phi}{f_{X}}+\beta_{n}\right).
\end{equation}

This periodicity is important for our cosmological considerations because this means $\phi$ cannot be arbitrarily large.
As we will see this essentially limits the maximal amount of dark matter we get.

In the following we will assume that one of the terms in Eq.~\eqref{goldpot} dominates. 
A suitable shift then also allows to remove the corresponding phase $\beta$.
Then we have a simple cosine structure,
\begin{equation}
V(\phi)=V_{n_{0}}\cos\left(n_{0} \frac{\phi}{f_{X}}\right).
\end{equation}
Typically we expect $n_{0}={\mathcal{O}}(\rm 1)$.

\section{Cold production from the misalignment mechanism}\label{misalignment}
Light scalar fields can be efficiently produced via the misalignment mechanism. In the following we will very briefly recall this mechanism~\cite{Preskill:1982cy,Abbott:1982af,Dine:1982ah} (we will follow~\cite{Arias:2012az}, see there for details), highlighting the constraints that will be relevant for our case.

In principle the spontaneous symmetry breaking giving rise to our pseudo-Goldstone boson can occur before or after inflation. In our discussion we will consider the former case. For a discussion of the latter case and its special features (which we expect to be very similar to the case of axion-like particles) see~\cite{Arias:2012az}.

Let us consider a constant initial value for the initial field (e.g. consider a situation after a period of inflation).
In the early Universe a real scalar (for small field values) evolves according to,
\begin{equation}
\ddot{\phi}+3H\dot{\phi}+m^{2}\phi=0.
\end{equation}

This equation, being equivalent to that of a damped harmonic oscillator has two regimes. For $H\gg m$ the oscillator is overdamped and the field remains essentially constant,
\begin{equation}
\phi\approx constant,\qquad H\gg m.
\end{equation}
This is the situation at very early times.
Accordingly at very early times there is no good reason why the field should be at its minimum. Since evolution is inhibited by the Hubble damping, there simply is no time for the field to evolve to its minimal value.

At later times when $H\ll m$ we get a weakly damped harmonic oscillator, and the field starts to perform oscillations with slowly decreasing amplitude. In the WKB approximation one obtains,
\begin{equation}
\phi(t)\simeq \phi_{1}\left(\frac{m_{1}a^{3}_{1}}{m(t) a^{3}(t)}\right)^{1/2}\cos\left(\int^{t}_{t_{1}} dt^{\prime} m(t^{\prime})\right).
\end{equation}
In this equation quantities with index $1$ are evaluated at the time when the field starts oscillating, 
\begin{equation}
3H(t_{1})=m(t_{1})=m_{1}.
\end{equation}
To be most general we allow for the mass to vary with time. This could, for example, be caused by thermal effects.
For axions this effect is quite important. 

For constant mass of the putative DM particle the energy density
\begin{equation}
\rho(t)\simeq \frac{1}{2}m(t)m_{1}\phi^{2}_{1}\left(\frac{a_{1}}{a(t)}\right)^{3}.
\end{equation}
is inversely proportional to the expanding volume, exactly what we expect for dark matter made from non-relativistic particles.

It is now convenient to translate scale factors into temperatures and to evaluate the energy density in $\phi$ today,
\begin{equation} 
\label{eq:CCDM}
\rho_{\phi,0}\simeq 0.17\, \frac{{\rm keV}}{{\rm cm}^3}\times \sqrt{\frac{m_0}{{\rm eV}}} \sqrt{\frac{m_0}{m_1}}\(\frac{\phi_1}{10^{11}\, {\rm GeV}}\)^2 {\cal F}(T_1).
\end{equation}
Here all quantities with index $0$ denote values today. 
${\mathcal F}(T_{1})=(g_{\star}(T_{1}/3.36)^{3/4}(g_{\star S}(T_{1})/3.91)^{-1}$ summarizes the dependence on the total number of degrees of freedom at the time when oscillations begin. It ranges from 1 to $\sim 0.3$ in the interval $T_{1}\in (T_{0},200\,{\rm GeV})$.

At this point it seems that we can always achieve the observed dark matter density,
\begin{equation}
\rho_{\rm CDM}=1.17\frac{\rm keV}{\rm cm^3},
\end{equation}
by choosing a suitable initial value $\phi_{1}$.
However, (pseudo-)Goldstone bosons do not allow for arbitrarily large field values. Indeed\footnote{At least for compact groups.} the maximal field value is $\sim f_{X}$.
We can use this to constrain the allowed regions for dark matter\footnote{In principle we could tune the initial value to be close to the maximum of the periodic pseudo-Goldstone potential. On the classical level being in this anharmonic region would modify Eq.~\eqref{eq:CCDM} by a function that diverges as we come closer and closer to this maximum. However, such a fine-tuning is difficult because of unavoidable fluctuations originating from the 
Hubble expansion as well as quantum effects. Nevertheless some increase in the dark matter density can be achieved~\cite{Wantz:2009it}.}.

\begin{figure}[t]
\centering
   \includegraphics[width=10cm]{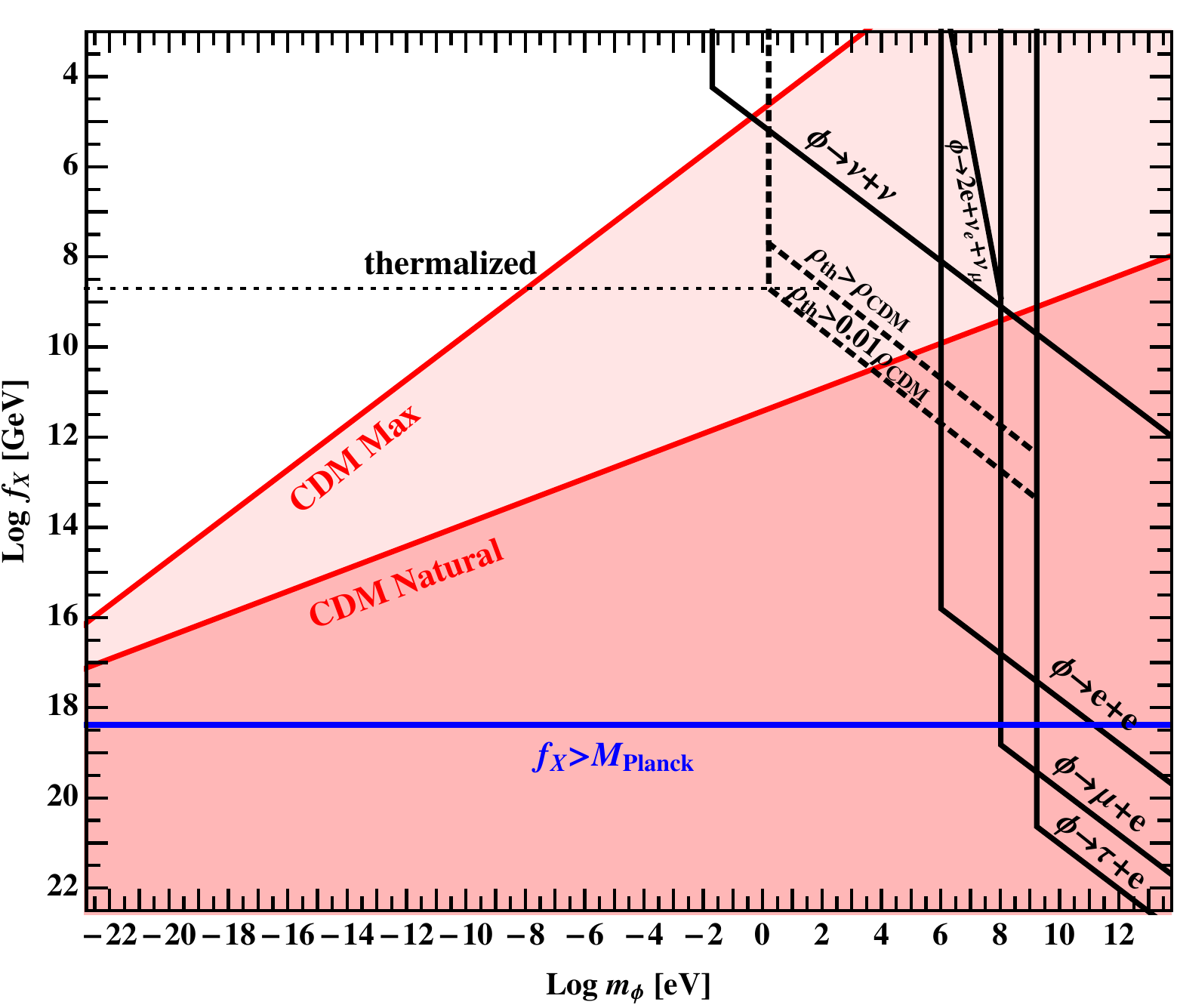} 
   \caption{Regions where pseudo-Goldstone particle produced from the misalignment mechanism can be dark matter (shaded red, for details see text). Constraints due to an insufficient lifetime from various possible decays (if possible) are shown as black lines. The thin dotted line is no exclusion but indicates the scale $f_{X}$ below which one expects a full thermal population of pseudo-Goldstone bosons (for a reheating temperature of $T_{R}=50\,{\rm GeV}$; it does not exist for very low reheating temperatures). The thick dashed lines indicate when the energy in this thermal population exceeds the given fraction of the cold dark matter density. The blue line indicates the Planck scale. We have used $\sqrt{a^2+b^2}\sim 1$.}
   \label{scalefig}
\end{figure}

We can now simply insert the maximal field value into Eq.~\eqref{eq:CCDM} to obtain viable dark matter regions for our pseudo-Goldstone particles.
Since we are interested in particles with very weak coupling to ordinary matter the simplest and perhaps most natural case seems to be a constant mass. In this case we find that for 
\begin{equation}
f_{X}\gtrsim 2.6\times 10^{11}\,{\rm GeV}\left(\frac{\rm eV}{m_{0}}\right)^{1/4}\left(\frac{1}{\mathcal{F}(T_{1})}\right)^{1/2},
\end{equation}
our particle can achieve a sufficient density to make up all of dark matter.
This is shown as the (stronger) red shaded area in Fig.~\ref{scalefig}.

However, a constant mass is not a necessary condition. Indeed, as the example of the axion shows, it is not even necessary for the new light particles to be in thermal equilibrium with ordinary matter for the mass to be changing quite dramatically\footnote{In the axion case the sector responsible for the generation of the mass, QCD, not the axions themselves are in thermal equilibrium. In general one could imagine similar things even occurring in hidden sector. If the latter is sufficiently colder than the visible sector problems with effective number of degrees of freedom can probably be avoided.}. So how much can the mass change? 
From the time of matter-radiation equality we are pretty sure that dark matter really behaves like non-relativistic particles. For our condensate this means that it must have started to oscillate. This corresponds to a lower limit on the mass~\cite{Arias:2012az,Das:2006ht}, 
\begin{equation}
m_{1}>3 H(T_{\rm eq})=1.8\times 10^{-27}\,{\rm eV}.
\end{equation}
Using Eq.~\eqref{eq:CCDM} this can be translated into a lower limit on $f_{X}$ for our pseudo-Goldstone bosons,
\begin{equation}
f_{X}\gtrsim 53\,{\rm TeV}\left(\frac{\rm eV}{m_{0}}\right)^{1/2}.
\end{equation}
This gives the larger light shaded region in Fig.~\ref{scalefig}.

As for the case of axions and axion-like particles one needs to take care that isocurvature fluctuations are not too large.
However, if we allow ourselves a sufficiently low reheating scale this typically does not impose very strong 
constraints (cf.~\cite{Arias:2012az}).

\subsection{Lifetime constraints}

\begin{figure}[t]
\centering
\subfigure[]{
\begin{picture}(180,100)(0,0)
\includegraphics[scale=0.4]{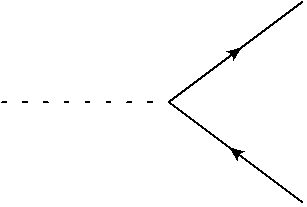}
\Text(-100,50)[c]{$\phi$}
\Text(-30,70)[c]{$\mu$}
\Text(-30,12)[c]{$\bar{e}$}
\end{picture}
\label{Yukawa}}
\subfigure[]{
\begin{picture}(250,100)(0,0)
\includegraphics[scale=0.4]{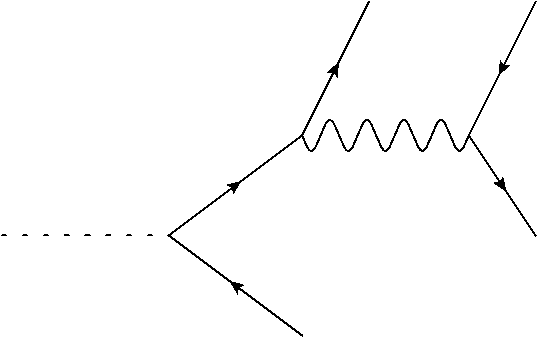}
\SetOffset(-93,0)
\Text(-100,50)[c]{$\phi$}
\Text(-30,70)[c]{$\mu$}
\Text(-30,12)[c]{$\bar{e}$}
\Text(0,110)[c]{$\nu_{\mu}$}
\Text(67,110)[c]{$\bar{\nu}_{e}$}
\Text(67,60)[c]{$e$}
\Text(35,67)[c]{$W^{-}$}
\end{picture}
\label{complicated}}
   \caption{Feynman diagrams for the decay of $\phi$ into SM fermions. We show the example of a coupling $\phi\mu e$.}
   \label{decaydiagram}
\end{figure}

The couplings of our light bosons to fermions can lead to decays. If it is kinematically allowed the most dangerous decay is the direct decay to these 
fermions as shown in Fig.~\ref{Yukawa} arising from interactions of the form Eqs.~\eqref{mix},\eqref{mux}.
The decay rate (ignoring small phase space corrections) is given by\footnote{See, e.g.~\cite{Cadamuro:2011fd} which also gives phase space corrections and comments on a thermal population.}
\begin{equation}
\Gamma(\phi\to f+h)
=\frac{m_{\phi}}{4\pi}(|Y_{fh}|^2+|Z_{fh}|^2),
\end{equation}
Using Eq.~\eqref{mux} we identify,
\begin{equation}
Y_{fh}=\frac{m_{f}-m_{h}}{f_{X}}a_{fh},\qquad Z_{fh}=\frac{m_{f}+m_{h}}{f_{X}}b_{fh} \end{equation}
this translates into
\begin{equation}
\Gamma(\phi\to f+h)=\frac{m_{\phi}}{4\pi}\frac{1}{f^{2}_{X}}\left[a^2_{fh}(m_{f}-m_{h})^{2}+b^2_{fh}(m_{f}+m_{h})^{2}\right].
\end{equation}
For $\phi$ to be a viable dark matter candidate the decay time must be smaller than the lifetime of the Universe. The limitations due to the above decays (if the corresponding coupling exists) are shown as the black lines in Fig.~\ref{scalefig}.

Below the mass threshold for these most simple decays, more complicated decays may still be possible.
If the coupling is flavor conserving, decay to photons or neutrinos may be possible. If it is flavor changing, decays with additional neutrinos may still occur, cf. Fig.~\ref{complicated}.

For $m_{\phi}\ll m_{\mu}$ the decay width for this decay (and its charge conjugate) is given by,
\begin{equation}
\Gamma(\phi\to 2e+2\nu)\sim \frac{G^{2}_{F}}{49152\,\pi^{5}}\frac{m^{7}_{X}}{f^{2}_{X}}\left[(a_{fh}-b_{fh})^2+corrections\right].
\end{equation}
(Similar for other couplings.) It is typically quite small.

\bigskip

In addition to direct decays of $\phi$ there is also the possibility for the condensate to evaporate via processes of the type
$\phi+f\rightarrow h+\gamma$. However, as discussed in~\cite{Arias:2012az} we expect this and similar processes to be suppressed for light pseudo-Goldstone bosons.

\subsection{Thermal population}
For sufficiently strong coupling we also expect a thermal population of $\phi$.
At temperatures above the mass of the fermions coupled to $\phi$ we expect interactions of the type
$h\rightarrow \phi f$ but also $h+\gamma\rightarrow f+\phi$.
For those\footnote{For non-condensate modes.} we expect interaction rates to be of order (assuming $\sqrt{a^2+b^2}\sim 1$ for simplicity of notation),
\begin{equation}
\label{rates}
\Gamma\sim \frac{1}{16\pi}\frac{m^{2}_{h}}{f^{2}_{X}} T,\quad{\rm and}\quad \Gamma\sim \alpha \frac{T^3}{f^{2}_{X}},
\end{equation}
where $h$ is the heavier of the two fermions coupled to $\phi$. 
For temperatures sufficiently above $m_{h}$ the second rate is larger. Therefore we will
focus in the following on this rate. 

If $\Gamma>H$ we expect that we have approximately an equilibrium number of $\phi$.
In the radiation dominated era we have,
\begin{equation}
H=\sqrt{\frac{\rho}{3 M^{2}_{P}}}\sim \sqrt{\frac{g_{\star}(T)}{10}}\frac{T^2}{M_{P}},
\end{equation}
where $g_{\star}(T)$ is the effective number of degrees of freedom.

$\Gamma/H$ increases with increasing temperature $T$. Therefore we expect the largest ratio at the highest temperatures.
Therefore we expect a thermal population of $\phi$ if
\begin{equation}
\label{thermalization}
f_{X}\lesssim 5\times 10^{8}\,{\rm GeV} \left(\frac{g_{\star}(T_{R})}{100}\right)^{1/4}\left(\frac{T_{R}}{50\,{\rm GeV}}\right)^{1/2},
\end{equation}
where $T_{R}$ is the reheating temperature.

However, having a thermal population of $\phi$ is not necessarily a problem. For small masses this simply behaves like an extra amount of relativistic energy, both at BBN as well as CMB release. Expressed as the effective number of neutrinos
this extra relativistic energy amounts to
\begin{equation}
\Delta N^{\rm \nu\, eff}_{\phi}=\left(\frac{g_{\star S}(T_{\rm dec})}{3.91}\right)^{-4/3}\frac{1}{\frac{7}{4}\left(\frac{4}{11}\right)^{4/3}}<0.6
\end{equation}
where the right hand side holds if $\phi$ decouples before the neutrinos. Moreover $T_{\rm dec}$ indicates the temperature at which $\phi$ decouples from the Standard Model particles.

Both BBN and CMB~\cite{Ade:2013zuv} observations are still compatible with $\Delta N_{\rm eff}\sim 0.6$.
Therefore with current limits this does not pose a constraint as long as there is only one $\phi$.
Turned around one may, of course speculate, that the slight indication for a non-vanishing $\Delta N_{\rm eff}$ (see~~\cite{Ade:2013zuv}) could be from such a thermal population of $\phi$. For masses below $\sim {\rm few}\times {\rm eV}$ (see below) this could be taken as a hint for a value $f_{X}\lesssim 5\times 10^{8} \,{\rm GeV}$.

For larger masses, however, the generated population will not just give extra relativistic energy, but it will give a hot dark matter component. For the case when $\phi$ is light decoupling happens when $\phi$ is relativistic.
The density of thermal $\phi$s then is,
\begin{equation}
\rho_{\rm th}\sim \frac{\zeta(3)}{\pi^2}\left(\frac{3.91}{g_{\star S}(T_{\rm dec})}\right) T^{3}_{\rm CMB} m_{\phi}
\sim 0.007\, \rho_{\rm CDM}\left(\frac{100}{g_{\star S}(T_{\rm dec})}\right)\left(\frac{m_{\phi}}{\rm eV}\right).
\end{equation}
This is severely constrained by observations. In particular recent Planck data~\cite{Ade:2013zuv}
suggests that such a hot component is less than about 1\% of the cold dark matter density (see~\cite{Archidiacono:2013cha} for a recent discussion considering axions which should behave very similarly to the case we are interested in). 

For even larger masses $m_{\phi}\gtrsim {\rm keV}$, $\phi$ starts to become cooler and behave more like warm or even cold dark matter. However, its density should definitely not exceed the measured dark matter abundance.

In both cases it is clear that $\phi$ should not thermalize. For larger masses even a fraction of the thermal population results in a too large density. Assuming incomplete thermalization, we can estimate the produced density as $\sim \Gamma/H$ times the full equilibrium density.
Requiring that the density, $\rho_{\rm th}$ of this thermal population is smaller than a part $\rho^{\rm max}_{\rm th}$ of observed dark matter density  imposes the limit,
\begin{equation}
f_{X}\gtrsim 4\times 10^{8}\,{\rm GeV}\left(\frac{\rho^{\rm max}_{\rm th}}{0.01\,\rho_{\rm CDM}}\right)^{1/2} \left(\frac{m_{\phi}}{\rm eV}\right)^{1/2}\left(\frac{100}{g_{\star}(T_{R})}\right)^{1/4}\left(\frac{100}{g_{\star S}(T_{R})}\right)^{1/2}\left(\frac{T_{R}}{50\,{\rm GeV}}\right)^{1/2}.
\end{equation}

This limit, does however, depend on the reheating temperature $T_{R}$. In Fig.~\ref{scalefig} we have assumed a fairly low value of about $T_{R}\sim 50\,{\rm GeV}$. However, many models allow for even lower reheating temperatures.
Indeed, if the reheating temperature is below the mass of the heavier of the fermion to which $\phi$ couples this
limit disappears completely\footnote{For temperatures close to the mass of the heavier fermion, the first 
rate in Eq.~\eqref{rates} may be larger than the second, resulting in a slightly lower $f_{X}$ required for thermalization.}.

\section{Non-dark matter constraints on pseudo-Goldstone couplings} \label{diagonal}
\subsection{Flavor diagonal couplings}
For flavor diagonal couplings there is a crucial difference between scalar and pseudoscalar couplings.
As we can see from our example given in Eqs.~\eqref{example1},\eqref{example12}, at tree-level and for on-shell fermions only the pseudoscalar part leads to direct Yukawa-like interactions.
This part is therefore much more severely constrained. In the following we will focus on such pseudo-scalar interactions
but we note and stress, that due to their significantly less constrained nature scalar interactions deserve to be more carefully investigated.

\subsubsection*{First generation couplings}
For masses below $10\,{\rm keV}$ the strongest constraints on pseudo-scalar couplings on electrons
arises from energy loss in red-giant stars~\cite{Raffelt:1994ry}. (Indeed there are some intriguing hints from the cooling of white dwarfs~\cite{Raffelt:1985nj,Isern:2008nt,Isern:2012ef,Isern:2010wz,Corsico:2012ki}).
Using Eq.~\eqref{example12} the limit on the pseudoscalar Yukawa coupling of
\begin{equation}
Y^{ee}_{p}\lesssim 3\times 10^{-13} \qquad{\rm for} \quad m_{\phi}\lesssim 10\,{\rm keV}
\end{equation}
can be translated into a limit
\begin{equation}
f^{ee}_{X}\gtrsim 3\times 10^{9}\,{\rm GeV}\qquad{\rm for} \quad m_{\phi}\lesssim 10\,{\rm keV}.
\end{equation}
Comparing with Fig.~\eqref{scalefig} most of the parameter region of interest for dark matter is still available, even taking into account these strong constraints,.

\subsubsection*{Second (and third) generation couplings}
Flavor diagonal couplings on second (or third) generation leptons are significantly less constrained.
On the electron the strongest constraints arise from astrophysical environments. There the constraints benefit from the enormous number of electrons present.
However, this is not the case for muons or taus, which are absent due to their short lifetime, as well as the high energy required to produce them.

For masses not much larger than the muon mass a sensitive test for psuedoscalar couplings to the muon is $(g-2)_{\mu}$.
Assuming a (very conservative) error of this measurement of the order of $\Delta (g-2)_{\mu}/2=4\times 10^{-9}$
one can obtain a limit on the pseudoscalar coupling
\begin{eqnarray}
Z^{\mu\mu}&\leq& 8\times 10^{-4} \qquad {\rm for}\quad m_{\phi}=0,
\\\nonumber
Z^{\mu\mu}&\leq& 1.3\times 10^{-3} \qquad {\rm for}\quad m_{\phi}\leq 2m_{\mu}.
\end{eqnarray}
Using an analog of Eq.~\eqref{example12} this translates into
\begin{equation}
f^{\mu\mu}_{X} \geq 160\,{\rm GeV} \qquad {\rm for}\quad m_{\phi}\leq 2m_{\mu}.
\end{equation}
which is quite a weak constraint not even touching the region suggested by the simplest models for dark matter.

For the third generation taus the constraints are even weaker.

\subsection{Flavor non-diagonal couplings}\label{non-diagonal}
Importantly, flavor non-diagonal couplings give non-vanishing tree level interactions both for scalar and for pseudoscalar pseudo-Goldstones.

As in the case of second and third generation couplings constraints from astrophysics are relatively weak. The reason is that at least one particle is from the second or third generation. Although we can start with an electron, astrophysical processes typically do not have enough energy to convert this electron into a muon. Therefore production of flavor non-diagonal pseudo-Goldstones will typically only occur at higher order in the coupling (e.g. the production of two (light) pseudo-Goldstones).

On the other hand laboratory measurements allow for very sensitive tests of flavor violating couplings.
Let us consider the most strongly constrained case of a coupling to electrons and muons as given in the example specified in Eqs.~\eqref{mix},~\eqref{mux}.

\begin{figure}[t]
\centering
   \includegraphics[width=10cm]{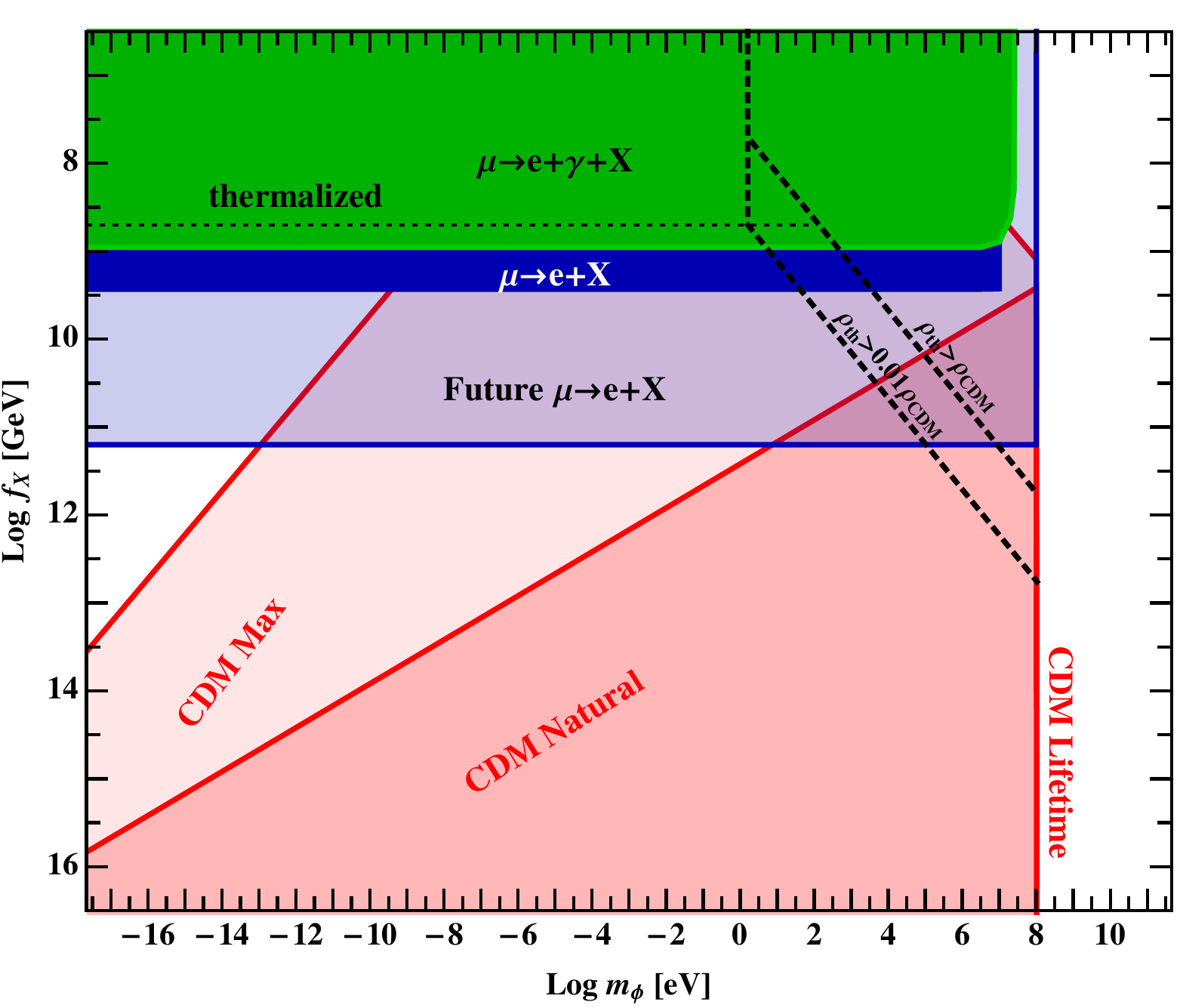} 
   \caption{Limits from various constraints on exotic muon decays. The blue shaded region roughly indicates the sensitivity of a possible future experiment~\cite{causality}. The red shaded areas give the region where a pseudo-familon could be a dark matter candidate. As in Fig.~\ref{scalefig} the dashed lines indicate the existence of a thermal population (thin) and where its density exceeds a fraction of the dark matter density (thick) if the reheating temperature is sufficiently high. For simplicity we have assumed $\sqrt{a^2+b^2}=a=1$.}
   \label{limitplot}
\end{figure}

Strong constraints arise in particular from searches for $\mu\to e+X$ and $\mu\to e+\gamma+X$, where $X$ specifies a single unobserved particle 
(not two neutrinos)~\cite{Jodidio:1986mz,Goldman:1987hy,Bolton:1988af,Feng:1997tn}, in our case the pseudo-Goldstone $\phi$.
The corresponding limits\footnote{For limits at higher scalar masses see~\cite{Harnik:2012pb}.} are indicated in Fig.~\ref{limitplot}.

\section{Conclusions}\label{conclusions}
In this note we have discussed the possibility that pseudo-Goldstone bosons, coupled to fermions via derivative interactions and produced via the vacuum misalignment mechanism can be a dark matter candidate. Family symmetries and the resulting familons are a prime example but our point is more general. 
If approximate family symmetries exist this seems a very natural option which is possible in quite a large region of parameter space; in particular also at quite low masses.

Although we have focussed mainly on a couple of examples in the lepton sector, the same reasoning can be applied to family symmetries in the quark sector. 
Both leptonic as well as quarkonic symmetries offer a wide range of possibilities for experimental tests by searching for the corresponding pseudo-Goldstone bosons in existing and future laboratory experiments.
 
Beyond that, the presence of a large number of these familons in the form of dark matter may also open new opportunities to test the origin of flavor experimentally.

\section*{Acknowledgements}
The author would like to thank Joachim Kopp, Javier Redondo, Andreas Ringwald and Andre Schoening for interesting discussions and helpful suggestions.

\end{document}